# Model of a Programmable Quantum Processing Device

Farid Ablayev[1], Sergey Andrianov[2], Danila Fetisov[1], Sergey Moiseev[3], Alexandr Terentyev[1], Andrey Urmanchev[1], and Alexander Vasiliev[1]

[1]Kazan (Volga Region) Federal University, Kazan, Russian Federation;
fablayev@gmail.com, fetisov.danila@gmail.com,
sapsan94@gmail.com, urmanchevandrey@gmail.com,
alexander.ksu@gmail.com
[2]Institute of Perspective Research, Kazan, Kazan, Russian Federation;
andrianovsn@mail.ru
[3]Kazan Quantum Center, Kazan, Russian Federation;
samoi@yandex.ru

**Abstract.** We propose a model of a programmable quantum processing device realizable with existing nanophotonic technologies and which can be viewed as a basis for new high performance hardware architectures. We present protocols and their physical implementation on the controlled photon transfer for executing basic single-qubit and multi-qubit gates. The possible operation of this quantum computer scheme is analyzed. The physical architecture is then formalized by a mathematical model of the Quantum Processing Unit (QPU), which is used as a basis for the Quantum Programming Framework that makes it possible to perform universal quantum computations in a multitasking environment.

**Keywords:** quantum computer · single-qubit and multi-qubit gates · decoherence · quantum processing unit · quantum memory

## 1 Introduction

Quantum computing provides a new paradigm for constructing high performance hardware architectures. There are many schemes for constructing quantum computer (QC). One of the key issues are the interconnections between computer elements [1]. They differ by their technological complexity, by the construction of multi-qubit quantum memory and its interface with quantum processers. This problem initiated the elaboration of so-called hybrid schemes for the universal QC due to large progress in recent studies of optical quantum memory (QM). Moreover, it was discovered that multi-qubit QM can provide an additional resource for quantum processing on photonic qubits [2].

Following this approach, recently we have proposed an extremely compact scheme of QC, where two blocks of multi-qubit QM being interconnected via gate atom placed in the same [3] or in two nearby QED cavities [4], [5].

Herein, it was proposed to accomplish CNOT operation using excitation transfer in this chain being driven at the gate atom by the control qubit preliminary extracted from the QM. These schemes seem to be compact and theoretically simple but it can be still challenging for implementation by using well-known nano-optical technologies. Therefore, in this paper we will develop new approach for implementing the excitation transfer based CNOT operations, where each atomic qubit is placed in its own cavity and all the cavities are interconnected directly or by nanofibers. For this purpose we study the whispering gallery mode resonators as good candidates for such cavities since they can possess sufficiently large Q-factor of about ($10^8$ -$10^{10}$) protecting qubits from the fast decoherence caused by the environment baths and can provide an efficient coupling of about 99.9% [6, 7] to the interconnecting nanofibers. In its turn, tapered fibers provide connection with external optical fiber link with efficiency up to 99.99% [8].

In this research we also propose an approach to make the future Quantum Processing Unit (QPU) programmable. We begin with formalization of the basic operations of such a device to show how it can perform an arbitrary quantum computation via universal basis set of gates. Then we present architecture of the Quantum Programming Framework (QPF) that operates the QPU via the classical controller and provides an application programming interface for running quantum computations in a classical multitasking environment.

## 2 Physical Model

Universal QC must contain a multi-qubit QM that stores the initial quantum information, the results of intermediate quantum operations and the final results of overall quantum computation. Logical quantum gates often consist of several elementary operations and are performed in several steps. Therefore, in addition to "operative" QM it is practical to have cache QM inserted in each separate quantum gate where the results of elementary operations are stored. This eliminates the necessity for repeated transfer of quantum information into external QM cells and vice versa, that can be subject to decoherence leading to loss of quantum information.

Let us consider the scheme of quantum transistor consisting of three micro-disks that are tunneling connected with each other. Disks *a*, *b* and *c* contain multi-atomic ensembles of three-level atoms that can be rare earth ions, NV-centers or quantum dots.

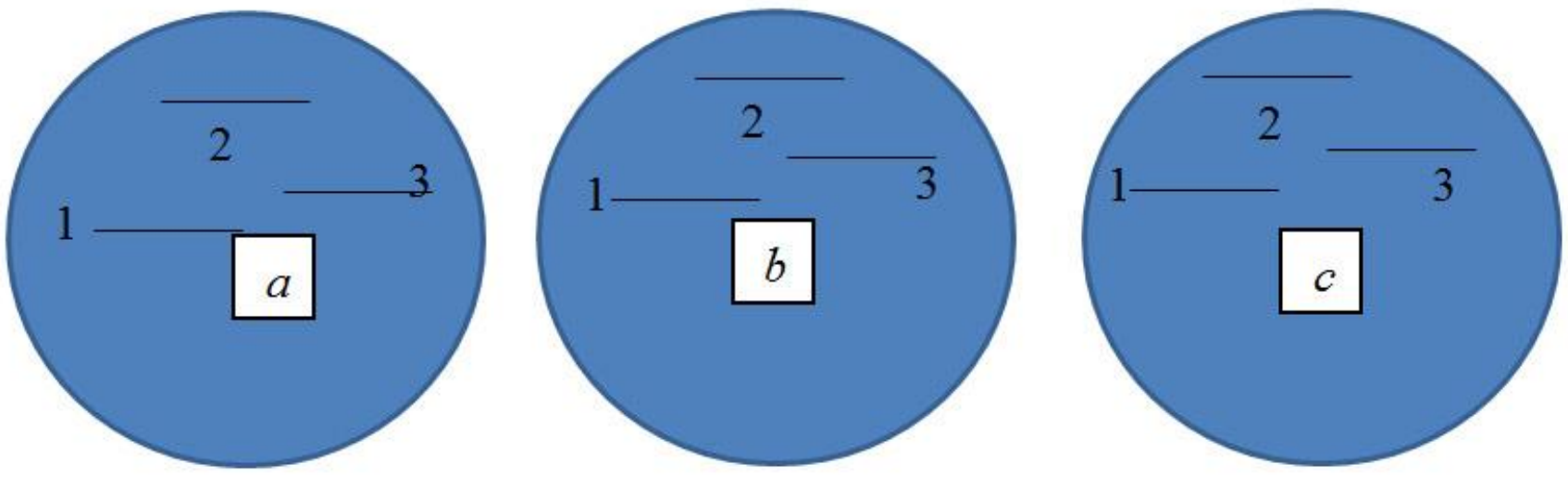

**Fig. 1.** Scheme of a quantum transistor.

## 2.1 Quantum Protocol

At the initial moment of time, logical qubit is loaded into the quantum memory of cavities *a* and *c* and three-level atom in the cavity *b* remains not excited if the control qubit was in the ground state and excited to the level 3 if the control qubit was in the excited state. As a result, the wave function of three cavities with quantum dots can be written as

$$\psi_1 = \alpha|0\rangle_{ph}^{(a)}|3\rangle_{mem}^{(a)}|0\rangle_{ph}^{(b)}|1\rangle_{dot}^{(b)}|0\rangle_{ph}^{(c)}|1\rangle_{mem}^{(c)} + \beta|0\rangle_{ph}^{(a)}|1\rangle_{mem}^{(a)}|0\rangle_{ph}^{(b)}|1\rangle_{dot}^{(b)}|0\rangle_{ph}^{(c)}|3\rangle_{mem}^{(c)} + \\ + \gamma|0\rangle_{ph}^{(a)}|3\rangle_{mem}^{(a)}|0\rangle_{ph}^{(b)}|3\rangle_{dot}^{(b)}|0\rangle_{ph}^{(c)}|1\rangle_{mem}^{(c)} + \delta|0\rangle_{ph}^{(a)}|1\rangle_{mem}^{(a)}|0\rangle_{ph}^{(b)}|3\rangle_{dot}^{(b)}|0\rangle_{ph}^{(c)}|3\rangle_{mem}^{(c)} \quad , \tag{1}$$

where $|n\rangle_{ph}^{(a,b,c)}$ is the wave function of cavities $a$, $b$, $c$ with the number of photons $n$ = 0,1; $|m\rangle_{mem}^{(a,c)}$ is the wave function of quantum memory (*a*,*c*) with the level number $m$ = 0,1,3; $|l\rangle_{dot}^{(b)}$ is the wave function of level $l$ = 1,2,3 of quantum dot in the cavity *b*.

Initially, excitation is transferred from quantum memory «*a*» into the cavity «*a*» by read out operation with induced transition from level 2 to level 3 and spontaneous transition $\hat{R}_{21}^{(a)}$ in the form of photon, afterwards quantum memory is switched off the cavity and the wave function takes the form

$$\psi_2 = \hat{R}_{12}^{(a)}\hat{U}_{23}^{(a)}\psi_1 = \alpha|1\rangle_{ph}^{(a)}|1\rangle_{mem}^{(a)}|0\rangle_{ph}^{(b)}|1\rangle_{dot}^{(b)}|0\rangle_{ph}^{(c)}|1\rangle_{mem}^{(c)} + \\ + \beta|0\rangle_{ph}^{(a)}|0\rangle_{mem}^{(a)}|0\rangle_{ph}^{(b)}|1\rangle_{dot}^{(b)}|0\rangle_{ph}^{(c)}|3\rangle_{mem}^{(c)} + \\ + \gamma|1\rangle_{ph}^{(a)}|1\rangle_{mem}^{(a)}|0\rangle_{ph}^{(b)}|3\rangle_{dot}^{(b)}|0\rangle_{ph}^{(c)}|1\rangle_{mem}^{(c)} + \\ + \delta|0\rangle_{ph}^{(a)}|1\rangle_{mem}^{(a)}|0\rangle_{ph}^{(b)}|3\rangle_{dot}^{(b)}|0\rangle_{ph}^{(c)}|3\rangle_{mem}^{(c)} \; . \tag{2}$$

Then, the interaction of the first and the second cavities is switched on, photon is transferred into the cavity «*b»* via quantum transfer operation $\hat{Q}_{ab}$ and the cavity «*a»* is switched off. Wave function takes the form

$$\begin{aligned}\psi_3 = \hat{Q}_{ab}\psi_2 = \alpha\left|0\right\rangle_{ph}^{(a)}\left|0\right\rangle_{mem}^{(a)}\left|1\right\rangle_{ph}^{(b)}\left|1\right\rangle_{dot}^{(b)}\left|0\right\rangle_{ph}^{(c)}\left|0\right\rangle_{mem}^{(c)} + \\ +\beta\left|0\right\rangle_{ph}^{(a)}\left|0\right\rangle_{mem}^{(a)}\left|0\right\rangle_{ph}^{(b)}\left|1\right\rangle_{dot}^{(b)}\left|0\right\rangle_{ph}^{(c)}\left|1\right\rangle_{mem}^{(c)} + \\ +\gamma\left|0\right\rangle_{ph}^{(a)}\left|0\right\rangle_{mem}^{(a)}\left|1\right\rangle_{ph}^{(b)}\left|3\right\rangle_{dot}^{(b)}\left|0\right\rangle_{ph}^{(c)}\left|0\right\rangle_{mem}^{(c)} + \\ +\delta\left|0\right\rangle_{ph}^{(a)}\left|0\right\rangle_{mem}^{(a)}\left|0\right\rangle_{ph}^{(b)}\left|3\right\rangle_{dot}^{(b)}\left|0\right\rangle_{ph}^{(c)}\left|1\right\rangle_{mem}^{(c)} ,\end{aligned} \tag{3}$$

Subsequently, we put the gate atom into the resonance and obtain:

$$\begin{aligned}\psi_4 = \hat{R}_{21}^{(b)}\psi_3 = \alpha\left|0\right\rangle_{ph}^{(a)}\left|0\right\rangle_{mem}^{(a)}\left|0\right\rangle_{ph}^{(b)}\left|2\right\rangle_{dot}^{(b)}\left|0\right\rangle_{ph}^{(c)}\left|0\right\rangle_{mem}^{(c)} + \\ +\beta\left|0\right\rangle_{ph}^{(a)}\left|0\right\rangle_{mem}^{(a)}\left|0\right\rangle_{ph}^{(b)}\left|1\right\rangle_{dot}^{(b)}\left|0\right\rangle_{ph}^{(c)}\left|1\right\rangle_{mem}^{(c)} + \\ +\gamma\left|0\right\rangle_{ph}^{(a)}\left|0\right\rangle_{mem}^{(a)}\left|1\right\rangle_{ph}^{(b)}\left|3\right\rangle_{dot}^{(b)}\left|0\right\rangle_{ph}^{(c)}\left|0\right\rangle_{mem}^{(c)} + \\ +\delta\left|0\right\rangle_{ph}^{(a)}\left|0\right\rangle_{mem}^{(a)}\left|0\right\rangle_{ph}^{(b)}\left|3\right\rangle_{dot}^{(b)}\left|0\right\rangle_{ph}^{(c)}\left|1\right\rangle_{mem}^{(c)} .\end{aligned} \tag{4}$$

Then, due to the interaction with $\pi$-pulse and withdrawing the gate atom from the resonance with cavity "*b*" we get the following state:

$$\begin{aligned}\psi_5 = R_{23}^{(b)}\psi_4 = \alpha\left|0\right\rangle_{ph}^{(a)}\left|0\right\rangle_{mem}^{(a)}\left|0\right\rangle_{ph}^{(b)}\left|3\right\rangle_{dot}^{(b)}\left|0\right\rangle_{ph}^{(c)}\left|0\right\rangle_{mem}^{(c)} + \\ +\beta\left|0\right\rangle_{ph}^{(a)}\left|0\right\rangle_{mem}^{(a)}\left|0\right\rangle_{ph}^{(b)}\left|1\right\rangle_{dot}^{(b)}\left|0\right\rangle_{ph}^{(c)}\left|1\right\rangle_{mem}^{(c)} + \\ +\gamma\left|0\right\rangle_{ph}^{(a)}\left|0\right\rangle_{mem}^{(a)}\left|1\right\rangle_{ph}^{(b)}\left|2\right\rangle_{dot}^{(b)}\left|0\right\rangle_{ph}^{(c)}\left|0\right\rangle_{mem}^{(c)} + \\ +\delta\left|0\right\rangle_{ph}^{(a)}\left|0\right\rangle_{mem}^{(a)}\left|0\right\rangle_{ph}^{(b)}\left|2\right\rangle_{dot}^{(b)}\left|0\right\rangle_{ph}^{(c)}\left|1\right\rangle_{mem}^{(c)} .\end{aligned} \tag{5}$$

Afterwards, we connect the cavities "*a*" and "*b*" again creating conditions for the transfer of photon from "*b*" to "*a*" with subsequent redistribution of excitation to level 3, so that we have the following state:

$$
\begin{aligned}
\psi_6 = U_{32}^{(b)} R_{21}^{(b)} Q_{ba} \psi_5 = &\alpha |0\rangle_{ph}^{(a)} |0\rangle_{mem}^{(a)} |0\rangle_{ph}^{(b)} |3\rangle_{dot}^{(b)} |0\rangle_{ph}^{(c)} |0\rangle_{mem}^{(c)} + \\
&+ \beta |0\rangle_{ph}^{(a)} |0\rangle_{mem}^{(a)} |0\rangle_{ph}^{(b)} |1\rangle_{dot}^{(b)} |0\rangle_{ph}^{(c)} |1\rangle_{mem}^{(c)} + \\
&+ \gamma |0\rangle_{ph}^{(a)} |3\rangle_{mem}^{(a)} |0\rangle_{ph}^{(b)} |2\rangle_{dot}^{(b)} |0\rangle_{ph}^{(c)} |0\rangle_{mem}^{(c)} + \\
&+ \delta |0\rangle_{ph}^{(a)} |0\rangle_{mem}^{(a)} |0\rangle_{ph}^{(b)} |2\rangle_{dot}^{(b)} |0\rangle_{ph}^{(c)} |1\rangle_{mem}^{(c)} .
\end{aligned}
\tag{6}
$$

Further, we convert the atomic excitation of "*c*" cavity quantum memory into the photon and perform the transfer of this excitation into the cavity "*b*". By (6) it gives:

$$
\begin{aligned}
\psi_7 = Q_{cb} \hat{R}_{12}^{(c)} U_{23}^{(c)} \psi_6 = &\alpha |0\rangle_{ph}^{(a)} |0\rangle_{mem}^{(a)} |0\rangle_{ph}^{(b)} |3\rangle_{dot}^{(b)} |0\rangle_{ph}^{(c)} |0\rangle_{mem}^{(c)} + \\
&+ \beta |0\rangle_{ph}^{(a)} |0\rangle_{mem}^{(a)} |1\rangle_{ph}^{(b)} |1\rangle_{dot}^{(b)} |0\rangle_{ph}^{(c)} |0\rangle_{mem}^{(c)} + \\
&+ \gamma |0\rangle_{ph}^{(a)} 3_{mem}^{(a)} |0\rangle_{ph}^{(b)} |2\rangle_{dot}^{(b)} |0\rangle_{ph}^{(c)} |0\rangle_{mem}^{(c)} + \\
&+ \delta |0\rangle_{ph}^{(a)} |0\rangle_{mem}^{(a)} |1\rangle_{ph}^{(b)} |2\rangle_{dot}^{(b)} |0\rangle_{ph}^{(c)} |0\rangle_{mem}^{(c)} .
\end{aligned}
\tag{7}
$$

After that, we act upon the gate atom that is out of resonance with cavity "*b*" by $\pi$-pulse at the transition $|3\rangle_{dot}^{(b)} \leftrightarrow |2\rangle_{dot}^{(b)}$ and get

$$
\begin{aligned}
\psi_8 = U_{23}^{(b)} \psi_7 = &\alpha |0\rangle_{ph}^{(a)} |0\rangle_{mem}^{(a)} |0\rangle_{ph}^{(b)} |2\rangle_{dot}^{(b)} |0\rangle_{ph}^{(c)} |0\rangle_{mem}^{(c)} + \\
&+ \beta |0\rangle_{ph}^{(a)} |0\rangle_{mem}^{(a)} |1\rangle_{ph}^{(b)} |1\rangle_{dot}^{(b)} |0\rangle_{ph}^{(c)} |0\rangle_{mem}^{(c)} + \\
&+ \gamma |0\rangle_{ph}^{(a)} |3\rangle_{mem}^{(a)} |0\rangle_{ph}^{(b)} |3\rangle_{dot}^{(b)} |0\rangle_{ph}^{(c)} |0\rangle_{mem}^{(c)} + \\
&+ \delta |0\rangle_{ph}^{(a)} |0\rangle_{mem}^{(a)} |1\rangle_{ph}^{(b)} |3\rangle_{dot}^{(b)} |0\rangle_{ph}^{(c)} |0\rangle_{mem}^{(c)} .
\end{aligned}
\tag{8}
$$

Thus, we have acted twice by $\pi$-pulse upon gate atom. Now, we introduce the gate atom into the resonance with the cavity "*b*" that leads to the following result:

$$
\begin{aligned}
\psi_9 = R_{21}^{(b)} \psi_8 = &\alpha |0\rangle_{ph}^{(a)} |0\rangle_{mem}^{(a)} |1\rangle_{ph}^{(b)} |1\rangle_{dot}^{(b)} |0\rangle_{ph}^{(c)} |0\rangle_{mem}^{(c)} + \\
&+ \beta |0\rangle_{ph}^{(a)} |0\rangle_{mem}^{(a)} |0\rangle_{ph}^{(b)} |2\rangle_{dot}^{(b)} |0\rangle_{ph}^{(c)} |0\rangle_{mem}^{(c)} + \\
&+ \gamma |0\rangle_{ph}^{(a)} |3\rangle_{mem}^{(a)} |0\rangle_{ph}^{(b)} |3\rangle_{dot}^{(b)} |0\rangle_{ph}^{(c)} |0\rangle_{mem}^{(c)} + \\
&+ \delta |0\rangle_{ph}^{(a)} |0\rangle_{mem}^{(a)} |1\rangle_{ph}^{(b)} |3\rangle_{dot}^{(b)} |0\rangle_{ph}^{(c)} |0\rangle_{mem}^{(c)} .
\end{aligned}
\tag{9}
$$

Afterwards, we put the gate atom out of resonance and connect cavities "*b*" and "*c*" transferring with that emerged photon into the memory atoms:

$$\begin{aligned}\psi_{10} = U_{32}^{(c)} R_{21}^{(c)} Q_{cb} \psi_9 = \alpha |0\rangle_{ph}^{(a)} |0\rangle_{mem}^{(a)} |0\rangle_{ph}^{(b)} |1\rangle_{dot}^{(b)} |0\rangle_{ph}^{(c)} |1\rangle_{mem}^{(c)} + \\ + \beta |0\rangle_{ph}^{(a)} |0\rangle_{mem}^{(a)} |0\rangle_{ph}^{(b)} |2\rangle_{dot}^{(b)} |0\rangle_{ph}^{(c)} |0\rangle_{mem}^{(c)} + \\ + \gamma |0\rangle_{ph}^{(a)} |3\rangle_{mem}^{(a)} |0\rangle_{ph}^{(b)} |3\rangle_{dot}^{(b)} |0\rangle_{ph}^{(c)} |0\rangle_{mem}^{(c)} + \\ + \delta |0\rangle_{ph}^{(a)} |0\rangle_{mem}^{(a)} |0\rangle_{ph}^{(b)} |3\rangle_{dot}^{(b)} |0\rangle_{ph}^{(c)} |1\rangle_{mem}^{(c)} .\end{aligned} \tag{10}$$

Further, we put gate atom into the resonance with cavity "*b*" and obtain:

$$\begin{aligned}\psi_{11} = R_{21}^{(b)} \psi_{10} = \alpha |0\rangle_{ph}^{(a)} |0\rangle_{mem}^{(a)} |0\rangle_{ph}^{(b)} |1\rangle_{dot}^{(b)} |0\rangle_{ph}^{(c)} |1\rangle_{mem}^{(c)} + \\ + \beta |0\rangle_{ph}^{(a)} |0\rangle_{mem}^{(a)} |1\rangle_{ph}^{(b)} |1\rangle_{dot}^{(b)} |0\rangle_{ph}^{(c)} |0\rangle_{mem}^{(c)} + \\ + \gamma |0\rangle_{ph}^{(a)} |3\rangle_{mem}^{(a)} |0\rangle_{ph}^{(b)} |3\rangle_{dot}^{(b)} |0\rangle_{ph}^{(c)} |0\rangle_{mem}^{(c)} + \\ \delta |0\rangle_{ph}^{(a)} |0\rangle_{mem}^{(a)} |0\rangle_{ph}^{(b)} |3\rangle_{dot}^{(b)} |0\rangle_{ph}^{(c)} |1\rangle_{mem}^{(c)} .\end{aligned} \tag{11}$$

Finally, we remove the gate atom from the resonance with the cavity "*b*" and tune the resonance among the cavities "*b*" and "*a*" and transferring the excitation into atoms we have the realization of the following operation:

$$\begin{aligned}\psi_{12} = U_{32}^{(a)} R_{21}^{(a)} Q_{ab} \psi_{11} = \alpha |0\rangle_{ph}^{(a)} |0\rangle_{mem}^{(a)} |0\rangle_{ph}^{(b)} |1\rangle_{dot}^{(b)} |0\rangle_{ph}^{(c)} |1\rangle_{mem}^{(c)} + \\ + \beta |0\rangle_{ph}^{(a)} |1\rangle_{mem}^{(a)} |0\rangle_{ph}^{(b)} |1\rangle_{dot}^{(b)} |0\rangle_{ph}^{(c)} |0\rangle_{mem}^{(c)} + \\ + \gamma |0\rangle_{ph}^{(a)} |1\rangle_{mem}^{(a)} |0\rangle_{ph}^{(b)} |3\rangle_{dot}^{(b)} |0\rangle_{ph}^{(c)} |0\rangle_{mem}^{(c)} + \\ \delta |0\rangle_{ph}^{(a)} |0\rangle_{mem}^{(a)} |0\rangle_{ph}^{(b)} |3\rangle_{dot}^{(b)} |0\rangle_{ph}^{(c)} |1\rangle_{mem}^{(c)} .\end{aligned} \tag{12}$$

Thus, an ability of performing Quantum Excitation Transfer (QET) operation is proved for the case when the gate atom is in the state 1 and absence of this operation if the atom was initially in the state 3, so, overall, this gives the controlled-QET (CQET) operation. It can be shown that the QET can be performed in a more general parameterized form QET($\theta$) that implements a partial excitation transfer.

To complete the universal set of quantum operations we discuss the operation PHASE($\theta$), that shifts the phase for some basis states and is implemented by Zeeman shift for the levels of one of memory units applying magnetic field to it. Therefore, the proposed scheme of quantum computer is capable of performing the complete set of operations using only three micro-cavities. Computer resources can be expanded using many triples of such micro-cavities.

## 2.2 Elementary Operations

In the case of $R_{2\alpha}^{(i)}$ $(\alpha = 1,3)$ operation, we have the following Hamiltonian:

$$H = \varepsilon_1 \sum_j S_{11}^{(j)} + \varepsilon_2 \sum_j S_{22}^{(j)} + \varepsilon_3 \sum_j S_{33}^{(j)} + \hbar\omega_\alpha a_\alpha^+ a_\alpha + \\ + \sum_j \left( g_{\alpha 2} a_\alpha^+ S_{\alpha 2}^{(j)} + g_{\alpha 2}^* a_\alpha S_{2\alpha}^{(j)} \right) \qquad (13)$$

with corresponding wave function

$$\psi = c_1 |0\rangle_a |1\rangle_{ph} + c_2 |1\rangle_a |0\rangle_{ph} = c_1\psi_1 + c_2\psi_2 \,, \qquad (14)$$

where $|0\rangle_a$ and $|1\rangle_a$ are wave functions corresponding to the ground state and to single photon excitation of multi-atomic ensemble, $|0\rangle_{ph}$ and $|1\rangle_{ph}$ are vacuum and single photon states.

Writing down the Schrodinger equation and finding the coefficients of wave function with respect to initial conditions $c_1(0) = \alpha$ , $c_2(0) = \beta$ , we get

$$c_1 = e^{-\frac{i}{2}(\omega_a + \omega_b)t} \alpha \cos\left\{ \sqrt{\frac{1}{4}(\omega_a - \omega_b)^2 + |\kappa|^2}\, t \right\} - \\ - ie^{-\frac{i}{2}(\omega_a + \omega_b)t} \frac{\alpha(\omega_a - \omega_b) + 2\beta\kappa}{\sqrt{(\omega_a - \omega_b)^2 + 4|\kappa|^2}} \sin\left\{ \sqrt{\frac{1}{4}(\omega_a - \omega_b)^2 + |\kappa|^2}\, t \right\}, \qquad (15)$$

$$c_2 = ie^{-\frac{i}{2}(\omega_a + \omega_b)t} \frac{\beta(\omega_a - \omega_b) + 2\alpha\kappa^*}{\sqrt{(\omega_a - \omega_b)^2 + 4|\kappa|^2}} \sin\left\{ \sqrt{\frac{1}{4}(\omega_a - \omega_b)^2 + |\kappa|^2}\, t \right\} + \\ + e^{-\frac{i}{2}(\omega_a + \omega_b)t} \beta \cos\left\{ \sqrt{\frac{1}{4}(\omega_a - \omega_b)^2 + |\kappa|^2}\, t \right\}, \qquad , (16)$$

where $\kappa = \frac{g_{\alpha 2}}{\hbar}\sqrt{N}$ , $\omega_a = \frac{\varepsilon_\alpha}{\hbar} + \omega_\alpha$ and $\omega_b = \frac{\varepsilon_2}{\hbar}$ . With $\omega_a = \omega_b$ , we have

$$c_1 = e^{-i\omega t}\{\alpha\cos(\kappa t) - i\beta\sin(\kappa t)\}, \quad (17)$$

$$c_2 = e^{-i\omega t}\{i\alpha\sin(\kappa t) + \beta\cos(\kappa t)\}. \quad (18)$$

It is obvious that when $\kappa t = \pi/2$, we have complete transfer of excitation.

Operation $Q_{ab}$ is governed by Hamiltonian

$$H = \hbar\omega_a a^+ a + \hbar\omega_b b^+ b + \hbar\kappa a^+ b + \hbar\kappa^* a b^+, \quad (19)$$

that yields a dynamics describing by the same formulas (15-18).

For the realization of PHASE($\theta$) gate, we first transform logical qubit into physical qubit situating on two atomic levels in one of cavities. It is described by wave function

$$\psi = c_1\psi_1 + c_2\psi_2, \quad (20)$$

where $\psi_1$ and $\psi_2$ are the states of spin $s = 1/2$. After the application of magnetic field *B* the quantum evolution is

$$\frac{dc_1}{dt} = -i\omega_a c_1, \quad (21)$$

$$\frac{dc_2}{dt} = -i\omega_b c_2, \quad (22)$$

where $\omega_a = \omega_0 + \frac{1}{2}g\mu B$ and $\omega_b = \omega_0 - \frac{1}{2}g\mu B$, $g$ is Lande factor, $\mu$ is Bor magneton.

The solutions of equations (21, 22) are

$$c_1 = C_1 e^{-i\omega_b t}, \quad (23)$$

$$c_2 = C_2 e^{-i\omega_b t}. \quad (24)$$

We see that the states of qubit acquire the mutual phase difference $\theta = g\mu Bt$.

# 3 Mathematical Model

A quantum processor with the above described architecture can be formalized by the following model.

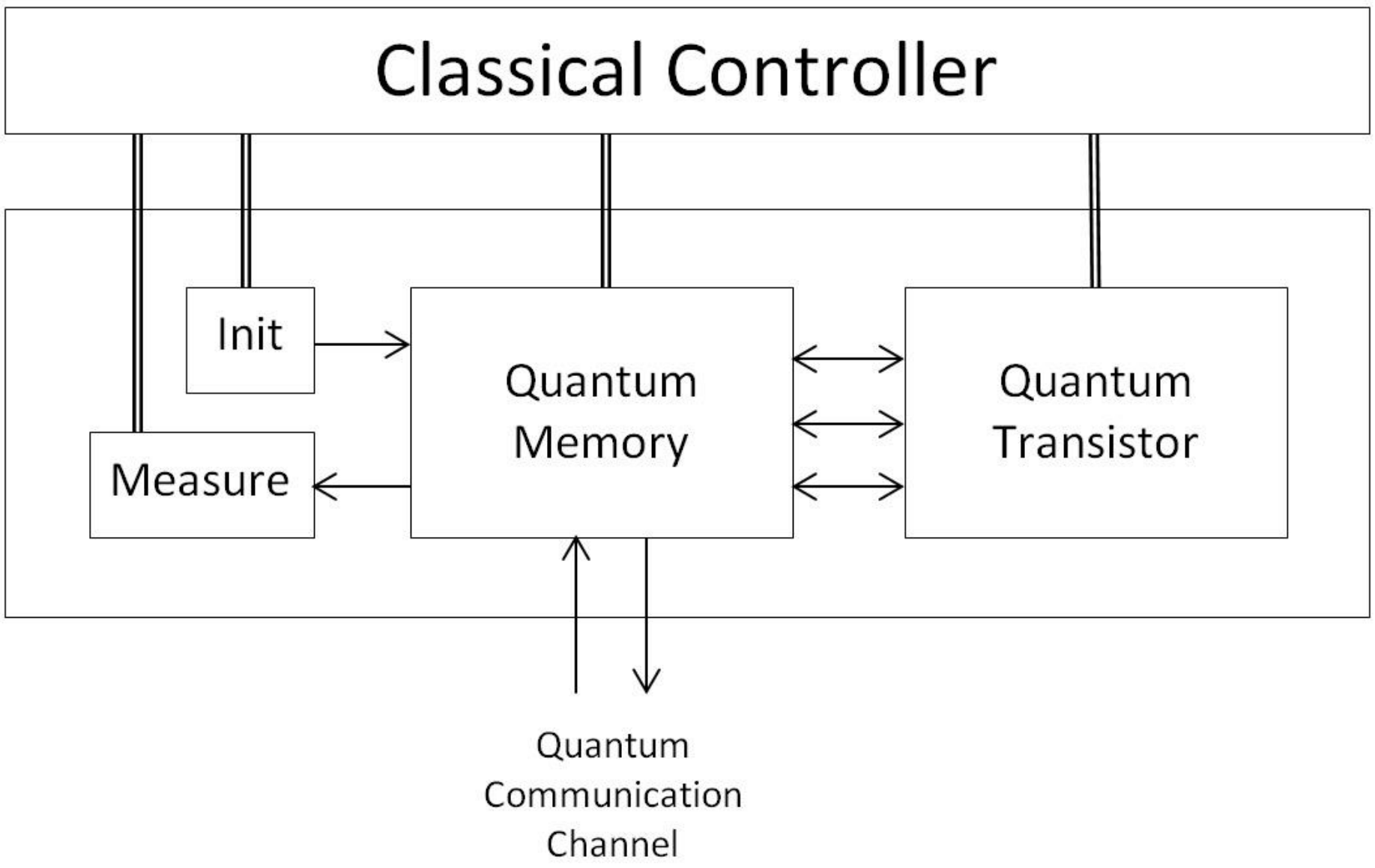

**Fig. 2.** A model of a Quantum Processing Unit (QPU).

A Quantum Processing Unit (QPU) is a device consisting of 4 modules controlled by a classical device as illustrated in Figure 1. They include:

- An $s$-qubit Quantum Memory module, that allows to store quantum states. The contents of the memory is described by the current state $|\psi\rangle \in \mathcal{H}^{2^s}$. The quantum memory module is capable of exchanging qubits with quantum transistor and quantum communication channel.
- A Quantum Transistor is a 3-qubit device capable of performing quantum transformations. There are three types of operations described in a physical architecture: QET, PHASE and CQET. Generally, we can consider an array of quantum transistors working in parallel, but here for simplicity we leave just one module.
- An Initialization Module can create a quantum state given its classical description and save it to the quantum memory. Without loss of generality we can consider a single-qubit initialization in state $|0\rangle$, since any other state can be obtained via universal set of operations.

- A Measurement Module implements a photon detecting procedure and gives a classical bit out of qubit being measured.

The proposed model is obviously a hardware-specific modification of the quantum circuit model of computation, and it is universal as well, since the available set of operations provides a universal basis for quantum computations [4].

A quantum algorithm in this model can be described by a $(t, s)$ *quantum program*, which is the sequence of $t$ instructions of several types (described below) over $s$-qubit quantum memory. These parameters correspond to the key complexity measures: $t$ is the time of computation and $s$ is the required space.

## 3.1 QPU Elementary Instructions

The set of elementary instructions correspond to the capabilities of module structure of the QPU and include the following operations.

1. INIT – an initialization command, that emits a single qubit in state $|0\rangle$, and saves it to quantum memory.

   Parameters: an address in the quantum memory

2. LOAD – transfers a specified qubit from the Quantum Memory to the Quantum Transistor.

   Parameters: an address in the quantum memory and the target transistor cell

3. SAVE – the reverse transfer from the Quantum Transistor to the Quantum Memory.

   Parameters: transistor cell number and the target address in the quantum memory

4. QET performs the following transformation on the pair of qubits loaded to the quantum transistor:

$$\mathrm{QET}(\theta) = \begin{pmatrix} 1 & 0 & 0 & 0 \\ 0 & \cos\frac{\theta}{2} & i\sin\frac{\theta}{2} & 0 \\ 0 & i\sin\frac{\theta}{2} & \cos\frac{\theta}{2} & 0 \\ 0 & 0 & 0 & 1 \end{pmatrix}$$

for arbitrary (experimentally possible) $\theta = \Omega_\sigma N t$.

Parameters: $\theta$ and the number of transistor (if there is more than one)

5. PHASE performs the following transformation on the pair of qubits loaded to the quantum transistor:

$$\mathrm{PHASE}(\theta) = \begin{pmatrix} 1 & 0 & 0 & 0 \\ 0 & e^{-i\theta/2+i\phi/2} & 0 & 0 \\ 0 & 0 & e^{i\theta/2+i\phi/2} & 0 \\ 0 & 0 & 0 & 1 \end{pmatrix},$$

where the phases $\theta$ and $\varphi$ are parameters of the physical implementation.

Parameters: $\theta$ and the number of transistor (if there is more than one).

6. CQET is a controlled version of the $\mathrm{QET}(\pi)$ operation which is implemented via quantum transistor effect as described in the previous section. This operation is given by the following unitary matrix:

$$\mathrm{CQET}(\theta) = \begin{pmatrix} 1 & 0 & 0 & 0 & 0 & 0 & 0 & 0 \\ 0 & \cos\frac{\theta}{2} & i\sin\frac{\theta}{2} & 0 & 0 & 0 & 0 & 0 \\ 0 & i\sin\frac{\theta}{2} & \cos\frac{\theta}{2} & 0 & 0 & 0 & 0 & 0 \\ 0 & 0 & 0 & 1 & 0 & 0 & 0 & 0 \\ 0 & 0 & 0 & 0 & 1 & 0 & 0 & 0 \\ 0 & 0 & 0 & 0 & 0 & 1 & 0 & 0 \\ 0 & 0 & 0 & 0 & 0 & 0 & 1 & 0 \\ 0 & 0 & 0 & 0 & 0 & 0 & 0 & 1 \end{pmatrix}$$

We let $\theta = \pi$.

Parameters: 1) the number of transistor (if there is more than one).

7. MEASURE operation extracts a qubit from the quantum memory and sends it to the Measurement Module and returns the results of the measurement.

Parameters: an address in the quantum memory.

Returns: the measurement result.

## 3.2 Logical Operations and Universality

We have shown earlier in [4], that the set of operations $\{\mathrm{QET}(\theta),\ \mathrm{PHASE}(\theta),\ \mathrm{CQET}\}$ is a universal one, which means that the presented model is capable of performing arbitrary quantum computations. The universality of this set is actually an encoded one [9], i.e. it is valid in the Hilbert subspace that corresponds to some logical encoding of qubits. Here we use the pairwise encoding mentioned in [10]: $|0_L\rangle = |01\rangle, |1_L\rangle = |10\rangle$.

In this encoding any single qubit state $\alpha|0_L\rangle + \beta|1_L\rangle$ is actually stored as an entangled two-qubit state $\alpha|01\rangle + \beta|10\rangle$, that is, the basis state of such a composite qubit is determined by the basis state of the first qubit of a pair.

In this encoding the QET$(\theta)$ gate corresponds to the rotation $R_x(\theta)$ around the $x$ axis of the Bloch sphere and the PHASE$(\theta)$ is a logical $R_z(\theta)$ operation. As shown in [4] CQET is a logical CNOT gate up to the relative phase, which can be made global by an additional PHASE$(\theta)$ operation.

## 4 Programming the QPU

In order to make a QPU programmable via the classical controller, we've developed a framework running over JAVA virtual machine. This framework has two main parts - the dispatcher and the service. The service adapts the high-level user commands to the low level for further implementation. The dispatcher was created to support the different types of connections between the classical computer and the QPU - it can be either a physical device or a QPU emulator.

Service is an intermediate layer between the client machine and the dispatcher. In order to handle client requests, transform and transfer the data to the controller the service is divided into the following layers:

- Code analysis layer
- Code transformation layer
- Code buffering layer

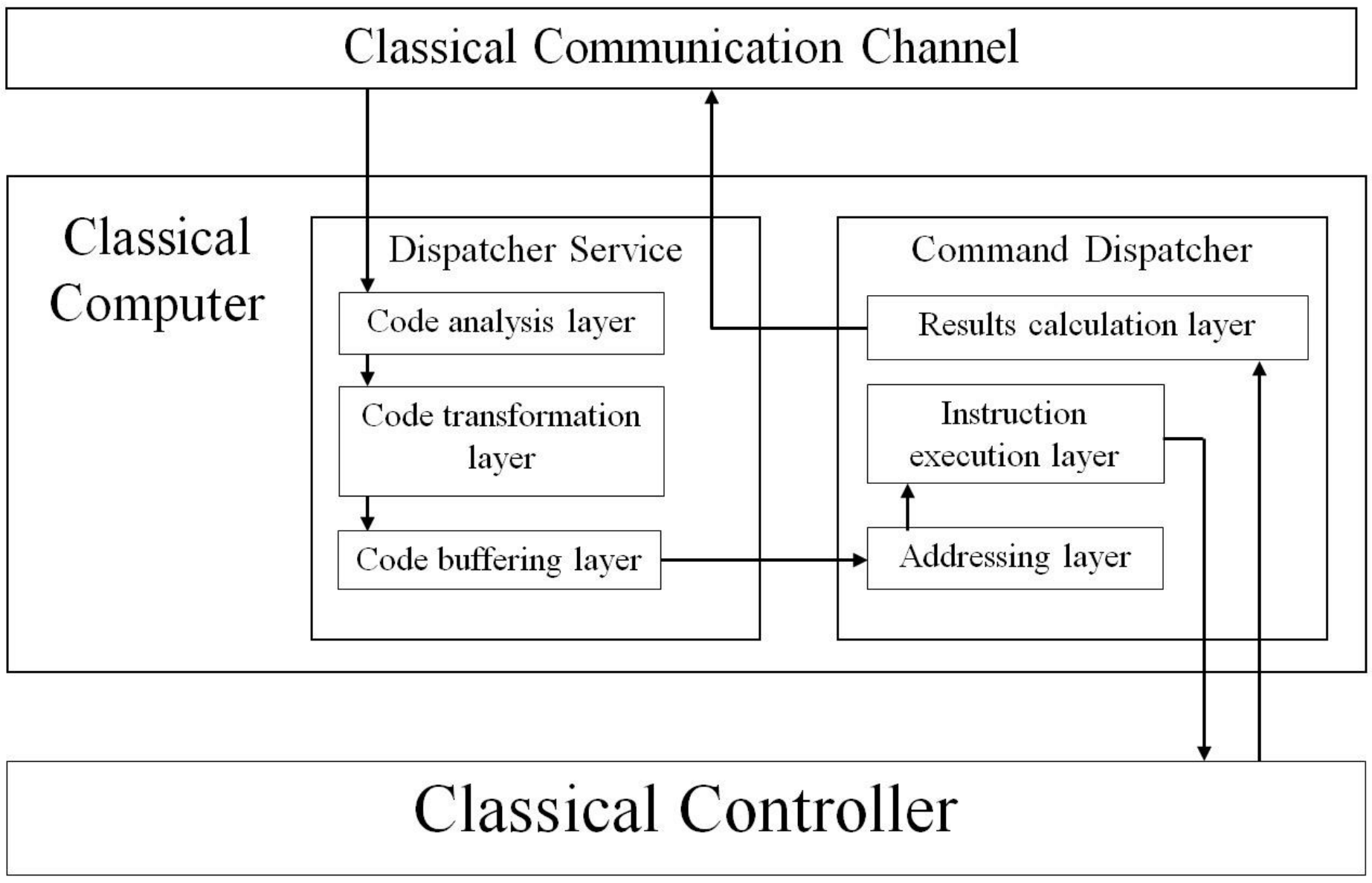

**Fig. 3.** The architecture of a Quantum Programming Framework.

The service can accept requests from multiple clients simultaneously. User interacts with the service by sending an array of data, which consist of operations from the universal basis $\{\mathrm{QET}(\theta),\ \mathrm{PHASE}(\theta),\ \mathrm{CQET}\}$, applied to the logical qubits, local addresses of logical qubits and operations parameters.

First of all, the data received from the user gets into the code analysis layer. This layer performs the following steps of code verification:

- Checking correctness of user's dataset.
- Checking availability of necessary parameters for operations from the universal basis
- Checking correctness of qubits addresses.

If the data successfully passes all steps, it will be transferred to the code transformation layer, otherwise the detailed description of an error will be returned to the client.

In the code transformation layer we perform the following steps:

- Assigning a global addresses to all logical qubits
- Dividing each logical qubit into two physical and transform operations

To facilitate the work of the dispatcher we decided to store a table of pairs - user's qubit local address and its global address in our service. We also store a correspondence table of logical qubit global address and global addresses of its two physical components. After splitting the logical qubit into two physical and assigning global addresses, we transform operations from user's data that work with logical qubits to operations that work with physical qubits.

Next, the data is passed to the code-buffering layer. After transformation of operations, we put them in the execution queue, which is built in following way: after we get the result from dispatcher, we request the maximal number of commands the controller can handle, create a sequence with the client's data from the waiting queue and send it all to dispatcher.

Dispatcher on its side pushes the command queue through the following processing layers:

- Address layer
- Command execution layer
- Result computation layer

In the first layer every qubit gets physical address parameters (index, frequency and recording time). Our aim is to decrease the downtime of the controller, so we maximally fill it with commands. To do that, we allow different logical qubits with same physical addresses in the same sequence. In this case system works correctly because in the beginning of the execution layer commands from different clients were split by the measure commands. This prevents different client's qubit sets from physically intersecting. Also, in second layer we insert initialization command before operations with qubits, which are not in the quantum memory yet, and submit the final sequence for execution.

In the third processing layer we accumulate measurement results received from the QPU, match them with logical qubit addresses and send back to the service.

The final result retrieved from dispatcher has the following structure: global address of the physical qubit and its measurement result. Using correspondence tables, which were discussed earlier, we transform the result of two physical qubits measurements to a single logical qubit measurement result. Here, we need to convert global address of the logical qubit to the client's local qubit address, collect qubit measurement results from each array in a separate object, and return to the client.

## 5 Conclusion

Thus, we have considered the operation of optical QC implementing basic operations on the controlled excitation transfer. Herein for the quantum controlling, we used three-level atoms in triples of whispering gallery resonators coupled with nanofibers. With that each coupled triples of cavities can serve both for processing and storing functions. Though here we have rather large number of steps for each quantum operation, each step can be performed with quite large efficiency using existing technologies and overall computer operation is productive bearing in mind that the storage of quantum information on nuclear sublevels of rare-earth ions can reach several hours [11].

The experimental implementation of the described QC architecture would require the integration of such a device in existing classical computing environments. We have proposed a Quantum Programming Framework based on the mathematical model of the Quantum Processing Unit that gives an application programming interface and allows to run quantum algorithms in a classical multitasking environment.

**Acknowledgements.** The work is performed according to the Russian Government Program of Competitive Growth of Kazan Federal University. Work was in part supported by the Russian Foundation for Basic Research (under the grants 14-07-00878, 15-37-21160).

## 6 References

1. Monroe, C.R., Schoelkopf, R.J., Lukin, M.D.: Quantum Connections and the Modular Quantum Computer. Sci. Am. 50 (2016).

2. DiCarlo, L., Chow, J.M., Gambetta, J.M., Bishop, L.S., Johnson, B.R., Schuster, D.I., Majer, J., Blais, A., Frunzio, L., Girvin, S.M., Schoelkopf, R.J.: Demonstration of two-qubit algorithms with a superconducting quantum processor. Nature. 460, 240–244 (2009).

3. Moiseev, S.A., Andrianov, S.N.: A quantum computer on the basis of an atomic quantum transistor with built-in quantum memory. Opt. Spectrosc. 121, 886–896 (2016).

4. Ablayev, F.M., Andrianov, S.N., Moiseev, S.A., Vasiliev, A. V: Quantum computer with atomic logical qubits encoded on macroscopic three-level systems in common quantum electrodynamic cavity. Lobachevskii J. Math. 34, 291–303 (2013).

5. Moiseev, S.A., Andrianov, S.N., Moiseev, E.S.: A quantum computer in the scheme of an atomic quantum transistor with logical encoding of qubits. Opt. Spectrosc. 115, 356–362 (2013).

6. Coillet, A., Henriet, R., Phan Huy, K., Jacquot, M., Furfaro, L., Balakireva, I., Larger, L., Chembo, Y.K.: Microwave photonics systems based on whispering-gallery-mode resonators. J. Vis. Exp. 1–10 (2013).

7. Kippenberg, T.J., Holzwarth, R., Diddams, S.A.: Microresonator-based optical frequency combs. Science. 332, 555–9 (2011).

8. Hoffman, J.E., Ravets, S., Grover, J.A., Solano, P., Kordell, P.R., Wong-Campos, J.D., Orozco, L.A., Rolston, S.L.: Ultrahigh transmission optical nanofibers. AIP Adv. 4, (2014).

9. Kempe, J., Bacon, D., DiVincenzo, D.P., Whaley, K.B.: Encoded Universality from a Single Physical Interaction. Quantum Comput. Inf. 1, 33–55 (2001).

10. Palma, G.M., Suominen, K.-A., Ekert, A.K.: Quantum computers and dissipation. Proc. R. Soc. London. Ser. A Math. Phys. Eng. Sci. 452, 567–584 (1996).

11. Zhong, M., Hedges, M.P., Ahlefeldt, R.L., Bartholomew, J.G., Beavan, S.E., Wittig, S.M., Longdell, J.J., Sellars, M.J.: Optically addressable nuclear spins in a solid with a six-hour coherence time. Nature. 517, 177–180 (2015).